\documentstyle[epsfig,aps]{revtex}
\begin{document}
 \draft
\title{Quantum logic between atoms inside a high Q optical cavity}
\author{L. You, X. X. Yi, and X. H. Su}
\address{School of Physics, Georgia Institute of Technology, Atlanta GA
30332, USA}
\date{\today}
\maketitle
\begin{abstract}
We propose a protocol for conditional quantum logic between
two 4-state atoms inside a high Q optical cavity.
The process detailed in this paper
utilizes a direct 4-photon 2-atom resonant
process and has the added advantage of commonly addressing
the two atoms when they are inside the high Q optical cavity.
\end{abstract}

\pacs{03.67.Lx, 32.80.Wr, 42.50.-p}

\narrowtext

Quantum information science has rapidly developed into a
major theme of modern research in recent
years. Various physical implementations have been
proposed for studying quantum communications and,
in particular, for controlling quantum logic operations
between individual qubits inside a high Q optical
cavity \cite{qc}. The strongly coupled cavity QED
system is unique as it is among the
selected few where coherent dynamics at the
level of a single quanta (electron, atom, photon, or phonon)
have already been observed. Furthermore,
photons represent one of the best choices for
quantum information distribution and communication,
the prospect of inter-converting quantum information
between light and matter as afforded in the cavity QED
system has led to imaginations of quantum information
networks in the future \cite{jeff1}.

The first proposal for quantum computing
with atoms inside a high Q optical cavity
appeared in 1995 \cite{zoller}, when
Pellizzari {\it et al.} discovered that conditional
logic between two atomic qubits can be achieved
with the use of the common cavity
mode as a quantum data-bus.
The protocol of Ref. \cite{zoller} was based
on adiabatic passage making use of dark state
structures in the combined atom-cavity system.
It is robust against both atomic and cavity decays
and also immune to noise from externally applied lasers.
While much progress
has been made on the experimental side \cite{ion,blatt,jeff,rempe,har},
the successfully implementation of this atom-cavity
protocol between two atoms has yet to be achieved.
The main difficulties
are:
1) precisely localizing each atomic motional wave packet
(this is needed to ensure the Lamb-Dicke limit);
2) obtaining a double $\Lambda$-type, 6-state,
level diagram for each atom; and
3) individually addressing each atom during
the dynamic gate operation when both atoms are inside the cavity.
Recent successes in combining an ion trap with a cavity
\cite{ion,blatt} and
in realizing trapped atoms inside a cavity
\cite{jeff,rempe,chapman}
have raised the hope that the first difficulty will be eliminated.
It is therefore highly desirable to develop protocols that
could potentially overcome the latter two limitations.
We note a similar scenario appeared with the
ion trap based quantum computing implementation \cite{zol},
where the development of the commonly addressing
protocol \cite{molmer} has lead to the first deterministic
generation of 4-atom maximally entangled state \cite{monroe}.

\begin{figure}
\includegraphics[width=3.in]{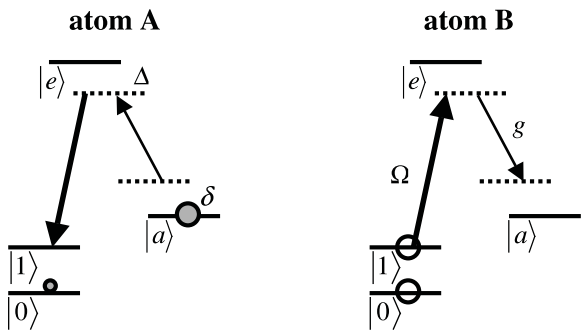}\\
\caption{Two 4-state atoms interacting with a
common cavity mode field.}
\label{fig1}
\end{figure}

In this paper, we suggest a protocol for quantum logic
between two 4-state atoms
that requires only common addressing
when atoms are inside the high Q optical cavity \cite{pachos}.
The system of two 4-level atoms inside a
single mode high Q optical cavity is described by
\begin{eqnarray}
H &&=H_A+H_B+H_C,\nonumber\\
H_{\mu} &&= \hbar\omega_1|1\rangle_\mu\langle 1|
+\hbar\omega_e|e\rangle_\mu\langle e|
+\hbar\omega_a|a\rangle_\mu\langle a|,\nonumber\\
&&+\left[{1\over 2}\hbar\Omega_\mu(t) e^{-i\omega_Lt} |e\rangle_\mu\langle 1|
+\hbar g_\mu(\vec r_\mu) c |e\rangle_\mu\langle a|+h.c.\right],\nonumber\\
H_C &&=\hbar\omega_C\, c^\dag c,
\end{eqnarray}
where $H_{\mu=A/B}$ and $H_C$ are respectively the Hamiltonian
for atom (A/B) and for the single mode cavity. The atomic
level scheme is as shown in Fig. \ref{fig1}.
$\Omega_\mu$ is the Rabi frequency due to an external laser
field ($|e\rangle\leftrightarrow|1\rangle$) at frequency $\omega_L$ and
$g_{\mu}$ (assumed real) is the single photon
coherent coupling ($|e\rangle\leftrightarrow|a\rangle$) rate with the cavity mode.
Similar models were considered earlier \cite{pachos,van,note}.

Before presenting the physical mechanism of our protocol,
we change to the interaction picture with
\begin{eqnarray}
U(t)&&=e^{-i(\omega_1+\omega_L)t\sum_{\mu}|e\rangle_\mu\langle e|
-i\omega_1t \sum_{\mu}|1\rangle_\mu\langle 1|}\nonumber\\
&&e^{-i\omega_at \sum_{\mu}|a\rangle_\mu\langle a|
-i[\omega_L-(\omega_a-\omega_1)]tc^\dag c}.
\end{eqnarray}
The resulting effective system dynamics
is then governed by \cite{how}
\begin{eqnarray}
{\cal H}_{\rm eff} &&=-\hbar\tilde\Delta_L
\sum_\mu|e\rangle_\mu\langle e|
-\hbar\tilde\delta_C\, c^\dag c\nonumber\\
&&+\sum_\mu\left[{1\over 2}\hbar\Omega_\mu(t) |e\rangle_\mu\langle 1|
+\hbar g_\mu c |e\rangle_\mu\langle a|+h.c.\right],
\end{eqnarray}
where we have defined
$\Delta_L=\omega_L-(\omega_e-\omega_1)$,
$\delta_C=(\omega_L-\omega_C)-(\omega_a-\omega_1)$,
and used the notations
$\tilde\Delta_L =\Delta_L+i{\Gamma/2}$,
$\tilde\delta_C =\delta_C+i\kappa$
that include the dissipative dynamics in their standard form.
$\Gamma$ denotes the atomic spontaneous emission rate
and $\kappa$ is the cavity decay rate (of each side).
We have neglected the position dependence
in the cavity coupling by assuming the Lamb-Dicke limit.

In our model, $|0\rangle$ and $|1\rangle$
are long lived atomic states, thus are
ideal candidates for atomic qubits.
The auxiliary atomic state $|a\rangle$
(also assumed to be long lived) is coupled
to the excited state $|e\rangle$
through the single mode cavity field.
Our protocol
starts with a Raman pulse (from two classical
laser fields) on atom A between
states $|1\rangle$ and $|a\rangle$ {\it via} $|e\rangle$.
After a time corresponds to a $\pi$ pulse
of the effective Raman interaction,
the state $|1\rangle_A$ is mapped onto $|a\rangle_A$.
Although this step requires the individual addressing
of atom A, it can be affected when the atom
is outside the tightly confined optical cavity.

The required conditional phase dynamics then read
\begin{eqnarray}
(\alpha|0\rangle+\beta|a\rangle)_A &&\otimes
(\mu|0\rangle+\nu|1\rangle)_B\otimes|0\rangle_C\nonumber\\
&&\to(\alpha|0\rangle+\beta|a\rangle)_A\otimes
\mu|0\rangle_B\otimes|0\rangle_C\nonumber\\
&&\ \,+(\alpha|0\rangle-\beta|a\rangle)_A\otimes
\nu|1\rangle_B\otimes|0\rangle_C,
\label{pg}
\end{eqnarray}
with the use of notation
$|i\rangle_A\otimes |j\rangle_B\otimes |k\rangle_C=|i,j,k\rangle$.
This conditional phase gate gives a relative phase
shift of $\pi$ between states
$|0,1,0\rangle$ and $|a,1,0\rangle$. Any
overall phase of the two states can be compensated
for by single bit operations on atom B \cite{univ}.

When atoms A and B are in the cavity, the
second step consists of using a common laser field
to drive atomic transitions
$|1\rangle\leftrightarrow|e\rangle$. The atom-atom
interaction required for conditional dynamics
(\ref{pg}) arises from the emission and absorption
of a common cavity photon. By operating in the
far off-resonant limit for all intermediate
one atom processes when $|\Delta_L|\gg \Omega_\mu|$,
$|\delta_C|\gg |g_\mu|$, and when the cavity is
initially empty (of photons), we realize an
effective 4-photon resonant process in the
following order $|a,1,0\rangle\to|a,e,0\rangle\to
|a,a,1\rangle\to|e,a,0\rangle\to|1,a,0\rangle$.
Upon elimination
of all intermediate states, we end up with an
effective 4-photon 2-atom coupling between
atomic states $|a,1\rangle$ and $|1,a\rangle$.
Naively we expect a 4-photon Rabi frequency being
$\propto {\Omega_A\Omega_B^*/ (4\Delta_L^2)}\times {g_Ag_B/\delta_C}$
when $|\Delta_L|\gg |{g_\mu^2/\delta_C}|$ is also satisfied.
Therefore, by driving a $2\pi$ pulse on this 4-photon
resonant transition, we gain a factor of ($-1$) in front of the
state $\beta\nu|a,1\rangle$.
As long as there is a large detuning $\delta_C$,
no other real transitions are possible.
Hence the conditional phase gate (\ref{pg}).
For atom B, its state $|1\rangle$ also
experiences a light shift of order ${|\Omega_B|^2/(4\Delta_L)}$
due to the classical field when $|\Delta_L|\gg |{g_\mu^2/ \delta_C}|$
is also satisfied.
The transition from $|1\rangle_B|0\rangle_C\to|a\rangle|1\rangle_C$
simply does not happen if the detuning $\Delta_L$ is made large.
The level shift becomes much more complicated if
$|\Delta_L|\gg |{g_\mu^2/\delta_C}|$ is not satisfied.

Now, we perform a detailed investigation
of the above envisioned 4-photon 2-atom transition.
We find the relevant dynamics by expanding the
state in the basis $|i,j,k\rangle$ with similarly
indexed coefficients $C's$ to arrive at
the following Schrodinger equation
\begin{eqnarray}
i\dot C_{1a0} &&={1\over 2}\Omega_A C_{ea0},\nonumber\\
i\dot C_{a10} &&={1\over 2}\Omega_B C_{ae0},\nonumber\\
i\dot C_{ea0} &&={1\over 2}\Omega_A^*C_{1a0}-\tilde\Delta_LC_{ea0}
+g_A C_{aa1}, \nonumber\\
i\dot C_{ae0} &&={1\over 2}\Omega_B^*C_{a10}-\tilde\Delta_LC_{ae0}
+ g_B C_{aa1},\nonumber\\
i\dot C_{aa1} &&=-\tilde\delta_CC_{aa1}
+g_AC_{ea0}+g_BC_{ae0}.
\end{eqnarray}
First we adiabatically eliminate the state $|a,a,1\rangle$
assuming $|\delta_C|\gg |g_\mu|$. Then the two states
$|e,a,0\rangle$ and $|a,e,0\rangle$ can be
eliminated as long as $|\Delta_L|\gg |\Omega_\mu|,|g_\mu^2/\delta_C|$
or $|\Delta_L-|g_\mu|^2/\delta_C|\gg |\Omega_\mu|.$
Finally we obtain the 4-photon resonant process between
states $|1,a,0\rangle$ and $|a,1,0\rangle$
\begin{eqnarray}
i\dot C_{1a0} &&=\delta\, C_{1a0}
+{1\over 2}\Omega_{\rm eff}C_{a10},\nonumber\\
i\dot C_{a10} &&={1\over 2}\Omega_{\rm eff}^*C_{1a0}
+\delta\, C_{a10},
\label{rb}
\end{eqnarray}
with
\begin{eqnarray}
\delta &&={1\over 2}{|\Omega|^2\over 4\tilde\Delta_L}
\left({1\over 1-2s}+1\right),\nonumber\\
\Omega_{\rm eff} &&={|\Omega|^2\over 4\tilde\Delta_L}
\left({1\over 1-2s}
-1\right).
\end{eqnarray}
$s={g^2/ (\tilde\Delta_L\tilde\delta_C)}$.
For simplicity we have assumed $g_A=g_B=g$
and $\Omega_A=\Omega_B=\Omega$ in the above
consistent with the common addressing requirement.
We note that an overall phase shift for the two states
given by
$\Theta(t)=\int_0^t\delta (t')dt'$ can be
simply absorbed to yield the solution to Eq. (\ref{rb})
\begin{eqnarray}
C_{1a0}(t)e^{i\Theta(t)} &&=C_{1a0}(0)\cos\theta(t)
-iC_{a10}(0)\sin\theta(t),\nonumber\\
C_{a10}(t)e^{i\Theta(t)} &&=C_{a10}(0)\cos\theta(t)
-iC_{1a0}(0)\sin\theta(t),
\end{eqnarray}
with the effective pulse area $\theta(t)=\int_0^t\Omega_{\rm eff} (t')dt'/2$.

The other group of coupled states is the single atom
off-resonant process on atom B. It can
be studied with
\begin{eqnarray}
|\psi(t)\rangle_I=C_{010}(t)|0,1,0\rangle+C_{0e0}|0,e,0\rangle
+C_{0a1}|0,a,1\rangle,\nonumber
\end{eqnarray}
and the corresponding Schrodinger equation
\begin{eqnarray}
i\dot C_{010} &&={1\over 2}\Omega_B C_{0e0},\nonumber\\
i\dot C_{0e0} &&={1\over 2}\Omega_B^*C_{010}-\tilde \Delta_LC_{0e0}+g_BC_{0a1},\nonumber\\
i\dot C_{0a1} &&=-\tilde\delta_CC_{0a1} +g_BC_{0e0}.
\end{eqnarray}
Following the same adiabatic elimination procedure
as used above, we arrive at
\begin{eqnarray}
i\dot C_{010} &&=\delta'\, C_{0e0},\nonumber\\
\delta' &&= {|\Omega|^2\over 4 \tilde\Delta_L}{1\over{1-s}},
\end{eqnarray}
i.e. a purely phase shift
$C_{010}(t)=C_{010}(0)e^{-i\Theta'(t)}$
with $\Theta'(t)=\int_0^t \delta'(t')dt'$.

In the limit when $|\Delta_L|\gg|\Omega|$, $|\delta_C|\gg |g|$,
and $|\Delta_L\delta_C|\gg g^2$, we find that
\begin{eqnarray}
\delta &&\approx
{|\Omega|^2\over 4\tilde\Delta_L}\left(1
+s\right)\approx \delta',\nonumber\\
\Omega_{\rm eff} &&\approx
{\Omega^2\over 2\tilde\Delta_L}s.
\end{eqnarray}
Although $|\Omega_{\rm eff}|\ll |\delta|,|\delta'|$,
the 4-photon Rabi frequency $\Omega_{\rm eff}$ does
affect a net ($-1$) phase shift upon completion of
its $(2\pi)$ pulse since $\Theta(t)\approx\Theta'(t)$.

\begin{figure}
\includegraphics[width=3.25in]{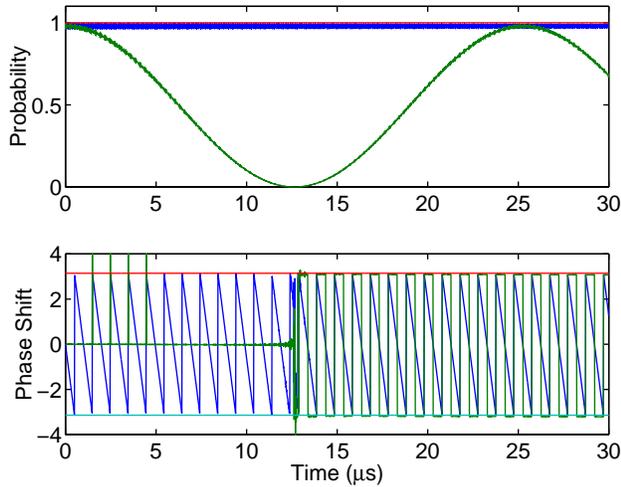}\\
\caption{Conditional phase dynamics
in the approximate adiabatic limit.}
\label{fig2}
\end{figure}

Figure \ref{fig2} shows a numerical simulation without
dissipations ($\Gamma=\kappa=0$). To satisfy the
adiabatic limit, we have used $\Omega=(2\pi)\,20$ (MHz),
$\Delta_L=(2\pi)\,100$ (MHz),
$g=(2\pi)\,10$ (MHz), and $\delta_C=(2\pi)\,50$ (MHz).
In the top panel, the oscillating line denotes
the probability amplitude of state $|a,1,0\rangle$,
which returns to its initial value upon completion
of the $2\pi$ pulse. The constant line
denote the probability of state $|0,1,0\rangle$
which remains at unity as expected.
The thickening of the lines is due to the rapid
secular oscillations.
In the lower panel, the saw-teeth like curve is the
absolute phase of the amplitude of state $|a,1,0\rangle$,
while the other solid line is for the relative
phase between states $|a,1,0\rangle$ and
$|0,1,0\rangle$. As expected it settles
down to $\pm\pi$.

\begin{figure}
\includegraphics[width=3.25in]{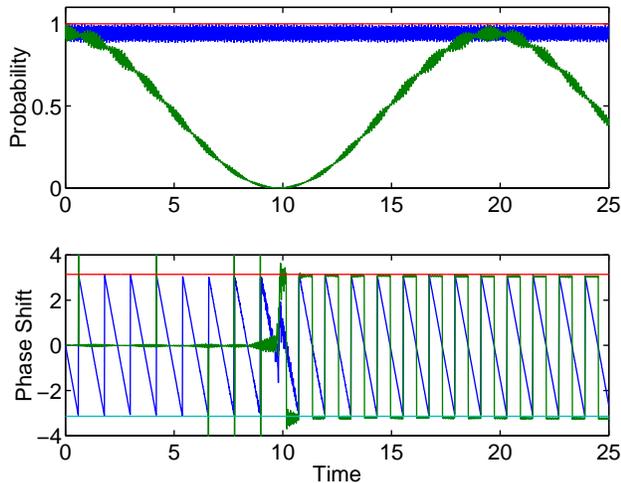}\\
\caption{The same as in Fig. 2, but not in the adiabatic limit.}
\label{fig3}
\end{figure}

Surprisingly, we find that the conditional dynamics
persists even beyond the limit when adiabatic elimination
is valid. For instance, in Fig. \ref{fig3} we display results
for $\Omega=(2\pi)\,10$ (MHz), $\Delta_L=(2\pi)\,30$ (MHz),
$g=(2\pi)\,3$ (MHz), and $\delta_C=(2\pi)\,8.75$ (MHz).
Apparently, the fact that both atoms share the same cavity
field data-bus is enough for establishing an effective interaction
between them.

Now we discuss effects of the dissipation/decoherence due to
both the cavity loss ($\kappa$) and the atomic decay ($\Gamma$).
As with any proposal for quantum computing implementation,
ultimately its success depends on being able to
complete many coherent
dynamics during the decoherence time.
In our case, as long as $|\Omega_{\rm eff}|\gg \Gamma,\kappa$,
we would expect essentially the same results as illustrated
in Figs. \ref{fig2} and \ref{fig3}.
On the other hand, this condition is difficult to
achieve because the 4-photon 2-atom resonant transition
is a relatively weak process due to large off-resosnant
detunings for all of its intermediate states.
In this respect, we find the second set of parameters
as used for Fig. \ref{fig3} more interesting as it
points to the use of longer cavities with
smaller $\kappa$ and $g$, as well as the use of atoms
with weaker transitions, thus smaller $\Gamma$.
During numerical simulations that include $\Gamma$
and $\kappa$, we find that in additional to a reduced success
rate and a slightly reduced fidelity of the logic gate,
the rapid secular oscillation is
also suppressed. Overall, it seems that atomic loss
$\Gamma$, rather than cavity decay $\kappa$ is the main
cause of failure as in the adiabatic passage protocol \cite{zoller}.
For instance,
with $\Gamma=(2\pi)0.03\, $(MHz) and $\kappa=(2\pi)0.1$(MHz),
the result for Fig. \ref{fig3}
remains essentially the same with a success rate close to 0.9.
For the parameters of Fig. \ref{fig2}, a success rate
larger than 0.9 is achieved when
$\Gamma=(2\pi)0.05\, $(MHz) and $\kappa=(2\pi)0.1$(MHz).

We note that there is wide regime of choices
for the external laser parameters $\Omega$ and $\Delta_L$.
In fact, similar outcomes are expected as long as their
ratio is maintained. This points to prospects of perhaps
using a higher order longitudinal
mode for the classical light such that it can also be
sent through the cavity directly \cite{jeff}.

Finally we want to stress that realizations of
all existing cavity QED based quantum logic protocols
remain challenging because of the technological limit
of the Fabry-Perot optical cavity. Realistically,
successful implementation of our protocol
requires $g^2\sim 10^3 \Gamma\kappa$, similar to
that required as in Ref. \cite{knight}, but
more stringent than Ref. \cite{pachos}
which requires $g^2\sim 10^2 \Gamma\kappa$. Using
the whispering gallery mode of a high Q optical
sphere, the parameter sets may be met if it is possible to
integrate with trapped atoms or ions \cite{hailing}.

In conclusion, we have suggested a new protocol
for conditional quantum phase gate between two atoms
inside a high Q cavity using a 4-photon 2-atom resonant
process. Our protocol eliminates
the difficult task of individual addressing of
atoms while they are inside the cavity and,
therefore, becomes easier to implement.
Furthermore, only 4-state atoms are used which
opens a wider opportunity of experimental choices.
Cavity QED based systems are usually deemed desirable,
because the possibility of converting quantum
information from atoms to photons for distribution
and communication, and because
of the potentially high clock cycle as afforded
in the strong coupling limit.
In the latter respect, similar to the recently
proposed {\it environment induced decoherence free space}
idea \cite{pachos,knight}
our protocol does not offer much advantage
as the inherent 4-photon 2-atom resonance at large
intermediate detunings results in relatively slow dynamics.

We thank Dr. M. S. Chapman for helpful comments.
This work is supported by a grant from NSA, ARDA, and DARPA
under ARO Contract No. DAAD19-01-1-0667, and by a grant
from the NSF PHY-0113831. Partial support from the NSF of
China is also acknowledged.


\begin{references}
\bibitem{note}In Ref. \cite{van}, a similar model
of two 4-state atoms was considered for a 2-photon
single atom resonant Raman coupling ($\delta_C=0$).
The conditional phase shift Eq. (\ref{pg}) can be
achieved by repeated operation on the single atom
of a Raman $\pi$-pulse in the order
(atom A)$\to$(atom B)$\to$(atom B)$\to$(atom A),
[rather than (atom A)$\to$(atom B)$\to$(atom A)$\to$(atom B)
as suggested in Ref. \cite{van}].
Our 4-photon protocol in a sense unites the 4 separate pulses
into one with the use of large off-resonant detunings for
all single atom processes. In doing so,
we end up with a protocol without the requirement
for individual atom addressing.
\end{references}
\end{document}